\documentstyle[12pt]{article}
\def\be {\begin{equation}}
\def\ee {\end{equation}}
\def\ba {\begin{eqnarray}}
\def\ea {\end{eqnarray}}

\def\bi {\begin{itemize}}
\def\ei {\end{itemize}}
\begin{document}
\def\bea{\begin{eqnarray}}
\def\eea{\end{eqnarray}}
\title{\bf {New Agegraphic Dark Energy in $f(R)$ Gravity}}
 \author{M.R. Setare  \footnote{E-mail: rezakord@ipm.ir}
  \\ {Department of Science,  Payame Noor University. Bijar, Iran}}
\date{\small{}}

\maketitle
\begin{abstract}
In this paper we study cosmological application of new agegraphic dark
energy density in the $f(R)$ gravity framework. We employ
 the new agegraphic model of dark energy to obtain the equation of state  for the new agegraphic energy density
 in spatially flat universe. Our calculation show, taking $n<0$ , it is possible to have $w_{\rm \Lambda}$ crossing
$-1$. This implies that one can generate phantom-like equation of
state from a new agegraphic dark energy model in flat universe in the
modified gravity cosmology framework. Also we develop a
reconstruction scheme for the modified gravity with $f(R)$ action.
 \end{abstract}

\newpage

\section{Introduction}
Many cosmological observations, such as SNe Ia [1], WMAP [2], SDSS
[3], Chandra X-ray observatory [4] etc., discover that our universe
is undergoing an accelerated expansion. They also suggest that our
universe is spatially flat, and consists of about
$70~^{\circ}/_\circ$ dark energy (DE) with negative pressure,
$30~^{\circ}/_\circ$ dust matter (cold dark matter plus baryons),
and negligible radiation. In order
to explain why the cosmic acceleration happens, many theories have
been proposed. It is the most accepted idea that a mysterious
dominant component, dark energy, with negative pressure, leads to
this cosmic acceleration, though its nature and cosmological origin
still remain enigmatic at present. An alternative proposal for dark
energy is the dynamical dark energy scenario. The cosmological
constant puzzles may be better interpreted by assuming that the
vacuum energy is canceled to exactly zero by some unknown mechanism
and introducing a dark energy component with a dynamically variable
equation of state. The dynamical dark energy proposal is often
realized by some scalar field mechanism which suggests that the
energy form with negative pressure is provided by a scalar field
evolving down a proper potential.\\
In recent years, many string theorists have devoted to understand
and shed light on the cosmological constant or dark energy within
the string framework. The famous Kachru-Kallosh-Linde-Trivedi (KKLT)
model \cite{kklt} is a typical example, which tries to construct
metastable de Sitter vacua in the light of type IIB string theory.
Furthermore, string landscape idea \cite{landscape} has been
proposed for shedding light on the cosmological constant problem
based upon the anthropic principle and multiverse speculation.
Although we are lacking a quantum gravity theory today, we still can
make some attempts to probe the nature of dark energy according to
some principles of quantum gravity. An interesting attempt in this
direction is the so-called ``holographic dark energy'' proposal
\cite{Cohen:1998zx,Hsu:2004ri,Li:2004rb,holoext}. Such a paradigm
has been constructed in the light of holographic principle of
quantum gravity  \cite{holoprin}, and thus it presents some
interesting features of an underlying theory of dark energy. More
recently a new dark energy model, dubbed agegraphic dark energy has
been proposed \cite{cai1} (see also \cite{zin}), which takes
 into account the Heisenberg uncertainty relation of quantum mechanics together with the gravitational
 effect in general relativity.\\
Because the holographic energy density belongs to a
dynamical cosmological constant, we need a dynamical frame to
accommodate it instead of general relativity. Einstein's theory of gravity may not describe gravity
at very high energy. The simplest alternative to general relativity
is Brans-Dicke scalar-tensor theory \cite{bd}. Modified gravity
provides the natural gravitational alternative for dark energy
\cite{odi1}. Moreover, modified gravity present natural unification
of the early-time inflation and late-time acceleration thanks to
different role of gravitational terms relevant at small and at large
curvature. Also modified gravity may naturally describe the
transition from non-phantom phase to phantom one without necessity
to introduce the exotic matter. But among the most popular modified
gravities which may successfully describe the cosmic speed-up is
$F(R)$ gravity. Very simple versions of such theory like $1/R$
\cite{1} and $1/R + R^2$ \cite{2} may lead to the effective
quintessence/phantom late-time universe (to see solar system
constraints on modified dark energy models refer to \cite{odin}).
Another theory proposed as gravitational dark energy is
scalar-Gauss-Bonnet gravity \cite{3} which is closely related with
low-energy string effective action.
\\
In present paper, using the new agegraphic model of dark energy in
spatially flat universe, we obtain equation of state for agegraphic
dark energy density in framework  of modified gravity. We show the phantomic description of the
new agegraphic dark energy in flat universe with $n<0$.
Also we develop a reconstruction scheme for the modified gravity
with $f(R)$ action, the known new agegraphic energy density is used for
this reconstruction.
\section{Modified gravity and new agegraphic dark energy}
 The action of modified gravity is given
 by
\begin{equation}
S=\int \sqrt{-g} d^{4}x[f(R)+L_{m}] . \label{action*}
\end{equation}
where $L_{m}$ is the matter Lagrangian density. The equivalent form
of above action is \cite{odi1}
\begin{equation}
S=\int d^{4}x \sqrt{-g}[P(\phi)R+Q(\phi)+L_{m}]. \label{action*1}
\end{equation}
where $P$ and $Q$ are proper functions of the scalar field $\phi$.
By the variation of the action (\ref{action*1}) with respect to the
$\phi$, we obtain
\begin{equation}
P'(\phi)R+Q'(\phi)=0 \label{action*2}
\end{equation}
which may be solved with respect to $\phi$:
\begin{equation}
\phi=\phi(R) \label{action*3}
\end{equation}
By the variation of the action (\ref{action*1}) with respect to the
metric $g_{\mu\nu}$, one can obtain
\begin{equation}
\frac{-1}{2}g_{\mu\nu}[P(\phi)R+Q(\phi)]-R_{\mu\nu}P(\phi)+\nabla_{\mu}\nabla_{\nu}P(\phi)-
g_{\mu\nu}\nabla^{2}P(\phi)+\frac{1}{2}T_{\mu\nu}=0
 \label{action*4}
\end{equation}
where $T_{\mu\nu}$ is the energy-momentum tensor. The equations
corresponding to standard spatially-flat FRW universe are
\begin{equation}
\rho=6H^2P(\phi)+Q(\phi)+6H\frac{dP(\phi)}{dt} \label{action*5}
\end{equation}
\begin{equation}
p=-(4\dot{H}+6H^2)P(\phi)-Q(\phi)-2\frac{d^{2}P(\phi)}{dt^2}-4H\frac{dP(\phi)}{dt}
 \label{action*6}
\end{equation}
where, $p$ and $\rho$ are the pressure and energy density due to the
scalar field in the modified gravity framework. By combining
(\ref{action*5}) and (\ref{action*6}) and deleting $Q(\phi)$, we
find
\begin{equation}
p+\rho=-2\frac{d^{2}P(\phi)}{dt^2}+2H\frac{dP(\phi)}{dt}-4\dot{H}P(\phi)
 \label{action*7}
\end{equation}
 Now we suggest a
correspondence between the new agegraphic dark energy scenario and the
above modified dark energy model. According to the new agegraphic dark energy we have following
relation for energy density \cite{cai}
 \be \label{holoda}
  \rho_\Lambda=3n^2M_{p}^{2}\eta^{-2}.
 \ee
 where the numerical factor $3n^2$ is introduced to parameterize some uncertainties, such as the species
of quantum fields in the universe, $\eta$ is conformal time, and
given by \be \label{con}
  \eta=\int\frac{dt}{a}=\int\frac{da}{a^2H}.
 \ee
 The critical energy density, $\rho_{cr}$, is given by following relation
\begin{eqnarray} \label{ro}
\rho_{cr}=3H^2
\end{eqnarray}
Using definitions
$\Omega_{\Lambda}=\frac{\rho_{\Lambda}}{\rho_{cr}}$ and
$\rho_{cr}=3M_{p}^{2}H^2$, we get

\begin{equation}\label{hl}
H\eta=\frac{n}{\sqrt{\Omega_{\Lambda}}}
\end{equation}
we obtain the equation of state for the agegraphic energy density.
Let us consider the dark energy dominated universe. In this case the
dark energy evolves according to  its conservation law \be
\dot{\rho}_{\Lambda}+3H(\rho_{\Lambda}+P_{\Lambda})=0
\label{coneq}\ee By considering  the definition of agegraphic
energy density $\rho_{\rm \Lambda}$, one
can find:
\begin{equation}\label{roeq}
\dot{\rho_{\Lambda}}=\frac{-2}{a\eta}\rho_{\Lambda}
\end{equation}
Substitute this relation into Eq.(\ref{coneq})  we obtain
\begin{equation}\label{stateq}
w_{\rm \Lambda}=\frac{2\sqrt{\Omega_{\Lambda}}}{3an}-1,
\end{equation}
then we can see that $w_{\Lambda}$ can cross the phantom divide if $n< 0$.
\\
As one can redefine the scalar field $\phi$ properly, we may choose
\begin{equation}\label{stateq1}
\phi=t.
\end{equation}
Now using Eqs.(\ref{holoda}), (\ref{stateq}), one can rewrite
Eq.(\ref{action*7}) as
\begin{equation}
2\frac{d^{2}P(t)}{dt^2}-2H\frac{dP(t)}{dt}+4\dot{H}P(t)-\frac{2\Omega_{\Lambda}^{3/2}H^2}{an}=0
 \label{action*8}
\end{equation}
In principle, by solving Eq.(\ref{action*8}) we find the form of
$P(\phi)$. Using Eqs. (\ref{action*5}), (\ref{holoda}), we also find
the form of $Q(\phi)$ as
\begin{equation}
Q(\phi)=3\Omega_{\Lambda}H^2-6H^2 P(\phi)-6H\frac{dP(\phi)}{dt}
 \label{action*9}
\end{equation}
\section{Modified gravity and its
 reconstruction from the new agegraphic dark energy}
In this section we consider another approach \cite{odi2} to
realistic cosmology in new agegraphic modified gravity. We start
with general $f(R)$-gravity action (\ref{action*}) but without the
matter term. For the spatially flat FRW universe we have
\begin{equation}
\rho=f(R)-6(\dot{H}+H^2-H\frac{d}{dt})f'(R)
 \label{action*10}
\end{equation}
\begin{equation}
p=f(R)-2(-\dot{H}-3H^2+ \frac{d^2}{dt^2}+2H\frac{d}{dt})f'(R)
 \label{action*11}
\end{equation}
where
\begin{equation}
R=6\dot{H}+12H^2
 \label{action*12}
\end{equation}
Again we use the new agegraphic dark energy density and substitute
Eq.(\ref{holoda}) into Eq.(\ref{action*10})
\begin{equation}
3\Omega_{\Lambda}H^2=f(R)-6(\dot{H}+H^2-H\frac{d}{dt})f'(R)
 \label{action*13}
\end{equation}
thus
\begin{equation}
f(R)=3\Omega_{\Lambda}H^2+6(\dot{H}+H^2-H\frac{d}{dt})f'(R)
 \label{action*14}
\end{equation}
Using Eqs.(\ref{holoda}), (\ref{stateq}), and substituting $f(R)$ into
Eq.(\ref{action*11}) one can obtain
\begin{equation}
\frac{d^2}{dt^2}f'(R)-H\frac{d}{dt}f'(R)+2\dot{H}f'(R)+f(R)+\frac{\Omega_{\Lambda}^{3/2}H^2}{an}=0
 \label{action*15}
\end{equation}
We shall consider the following simple solution
\begin{equation}\label{112}
a=a_0(t_s-t)^{h_0},
\end{equation}
where   $a_0$, $h_0$ and  $t_{s}$ are constant. Substituting
Eq.(\ref{112}) into Eq.(\ref{action*12}), give us following relation
for scalar curvature
\begin{equation}
R=\frac{12h_{0}^{2}-6h_0}{(t_s-t)^{2}}
 \label{action*122}
\end{equation}
Using Eqs.(\ref{con}, \ref{112}) we can write
\begin{equation}\label{116}
\eta=\int_{t}^{t_s}\frac{dt}{a_0(t_s-t)^{h_0}}=\frac{1}{a_0(1-h_0)(t_s-t)^{h_0-1}}
\end{equation}
Now using definition $\rho_{\Lambda}$ and above relation we obtain
the time behaviour of agegraphic dark energy as
\begin{equation}\label{117}
\rho_{\Lambda}=\frac{3n^2a_0^{2}(1-h_0)^{2}}{(t_s-t)^{2-2h_0}}
\end{equation}
Substituting the above $\rho_{\Lambda}$ into Eq.(\ref{action*10}),
and using Eqs.(\ref{112},\ref{action*122}) one can obtain
\begin{equation}\label{118}
\frac{72h_{0}^{2}(1-2h_0)}{(t_s-t)^{4}}f''(R)-\frac{6h_0(h_0-1)}{(t_s-t)^{2}}f'(R)+f(R)=
\frac{3n^2a_0^{2}(1-h_0)^{2}}{(t_s-t)^{2-2h_0}}
\end{equation}
Again we use Eq.(\ref{action*122}) and rewrite the above
differential equation as following
\begin{equation}\label{119}
f''(R)+\frac{a}{R}f'(R)+\frac{b}{R^2}f(R)= \frac{d}{R^{1+h_0}}
\end{equation}
where
\begin{equation}\label{120}
a=\frac{h_0-1}{2}, \hspace{1cm} b=\frac{1-2h_0}{2}, \hspace{1cm}
d=\frac{-a_0^{2}n^2(1-h_0)^{2}}{4}(6h_0(2h_0-1))^{h_0}.
\end{equation}
The solution of differential equation (\ref{119}) is given by
\begin{equation}\label{121}
f(R)=C_{1}R^{\frac{1}{2}\left(\frac{3-h_0}{2}-\sqrt{\frac{(h_0-3)^{2}}{4}+4h_0-2}\right)}+
C_{2}R^{\frac{1}{2}\left(\frac{3-h_0}{2}+\sqrt{\frac{(h_0-3)^{2}}{4}+4h_0-2}\right)}+\frac{n^2a_0^{2}(1-h_0)^{2}(6h_0(2h_0-1))^{h_0}}{2h_{0}(2-h_0)R^{h_0-1}}
\end{equation}
where $C_{1},C_{2}$ are constant. Therefore, a consistent modified
gravity whit new agegraphic dark energy in flat space has the above
form. In order that the accelerating expansion in the present
universe could be generated, let us consider that $f(R)$ could be a
small constant at present universe, that is,
\begin{equation}\label{123}
f(R_0)=-2R_0, \hspace{1cm} f'(R_0)\sim 0,
\end{equation}
where $R_0\sim (10^{-33}eV)^{2}$ is current curvature \cite{odin}.
By impose the conditions (\ref{123}) on the solution (\ref{121}) we
can obtain the constants $C_{1}$ and $C_{2}$ as following
\begin{equation}\label{124}
C_1=-\frac{(1-h_0)^{2}a_0^{2}n^{2}(6h_0(2h_0-1))^{h_0}(v+h_0-1)}{(v-uR_0)R_0^{h_0+u-1}}-\frac{2v}{(v-uR_0)R_0^{u-1}},
\end{equation}
\begin{equation}\label{125}
C_2=\frac{(1-h_0)^{2}a_0^{2}n^{2}(6h_0(2h_0-1))^{h_0}(uR_0+h_0-1)}{(v-uR_0)R_0^{h_0+v-1}}+\frac{2u}{(v-uR_0)R_0^{v-2}},
\end{equation}
where
\begin{equation}\label{126}
u=\frac{1}{2}\left(\frac{3-h_0}{2}-\sqrt{\frac{(h_0-3)^{2}}{4}+4h_0-2}\right),
\hspace{0.3 cm}
v=\frac{1}{2}\left(\frac{3-h_0}{2}+\sqrt{\frac{(h_0-3)^{2}}{4}+4h_0-2}\right)
\end{equation}

\section{Conclusions}
In order to solve cosmological problems and because the lack of our
knowledge, for instance to determine what could be the best
candidate for DE to explain the accelerated expansion of universe,
the cosmologists try to approach to best results as precise as they
can by considering all the possibilities they have. Within the
different candidates to play the role of the dark energy, the
new agegraphic dark energy model, has emerged as a possible model  with EoS across
$-1$. In the present paper we have studied
cosmological application of new agegraphic dark energy density in the $f(R)$
modified gravity framework. By considering the agegraphic energy
density as a dynamical cosmological constant, we have obtained the equation of state for the agegraphic energy density in the $f(R)$
gravity framework. We have shown if $n< 0$,
the new agegraphic dark energy model also will behave like a phantom
model of dark energy the amazing feature of which is that the
equation of state of dark energy component $w_{\rm \Lambda}$ crosses
$-1$. Also we have developed a
reconstruction scheme for modified gravity with $f(R)$ action. We
have considered the energy density in Eq.(\ref{action*10}) in
new agegraphic form, then by assumption a simple solution as
Eq.(\ref{112}) we could obtain a differential equation for $f(R)$,
the solution of this differential equation give us a modified
gravity action which is consistent with new agegraphic dark energy
scenario.

\end{document}